\documentclass[onecolumn,pra,aps,showpacs,reprint]{revtex4-1} 
\usepackage{amsmath,mathrsfs,amsbsy,color,graphicx,bm,amsthm,amsfonts}
\usepackage{times}
\usepackage{graphicx}
\usepackage{braket}
\usepackage{dsfont}

\newtheorem*{teo*}{Theorem}

\newtheorem{coro}{Corollary}

\newtheorem*{lema*}{Lemma}
\newtheorem{lema}{Lemma}
\DeclareMathOperator{\Tr}{Tr}

\begin{document}
\title{Quantifying resources for Page-Wootters mechanism: Shared asymmetry as relative entropy of entanglement}
\author{Rafael S. Carmo }
\email{rafael.carmo@gmail.com}
\author{Diogo O. Soares-Pinto}
\email{dosp@usp.br}
\affiliation{Instituto de F\'isica de S\~ao Carlos, Universidade de S\~ao Paulo, CP 369, 13560-970, S\~ao Carlos, S\~ao Paulo, Brazil}

\begin{abstract}
Recently, some attention has been given to the so-called Page-Wootters mechanism of quantum clocks. Among the various proposals to explore the mechanism using more modern techniques, some have chosen to use a quantum information perspective, defining and using informational measures to quantify how well a quantum system can stand as a reference frame for other quantum system. In this work, we explore the proposal based on resource theory of asymmetry, known as mutual or shared asymmetry, which actually is equivalent to the approach from coherence theory in the case of interest here: quantum reference frames described by the $U(1)$ compact group. We extend some previous results in literature about shared asymmetry and Page-Wootters mechanism to more general cases, culminating in the enunciation of a theorem relating shared asymmetry of a bipartite state $\rho_{SR}$ with the relative entropy of entanglement of \textit{internal states} $\rho_M$ on the charge sectors of the Hilbert space $\mathcal{H}_S\otimes\mathcal{H}_R$. Using this result we reinterpret the relation between Page-Wootters mechanism and entanglement and also open some paths to further studies.

\end{abstract}

\maketitle

\section{Introduction}
Physical systems are always described based on some reference frame that in general constitutes in an external classical system that appears just as a parameter in the equations. Two obvious examples in quantum theory are spacial and time reference frames which in the wave function under position representation, $\psi(\overrightarrow{x},t)$, is characterized by the parameters $\overrightarrow{x}$ and $t$. However, what happens if you eliminate these external reference frames and try to describe the states in a relational way, that is, using one quantum state as a reference for another?

Answers to this question have recently been given by the theory of quantum reference frames \cite{ReviewQuantumRef}. Using modern techniques like resource theory, \cite{teoriasderecurso}, asymmetry theory \cite{marvian2012}, coherence theory \cite{bcp2014, marvianspekkens2016} and quantum communication protocols \cite{kitaevmayerspreskill2004}, several studies have been developed with the objective of not only circumventing the practical difficulty of working with quantum states in the absence of an adequate external reference frames, but also to investigate more fundamentally how quantum mechanics could be used universally, without the need for privileged external reference frames and classical systems \cite{gambinipullin2015, poulin2006}.

Mathematically, the absence of a reference frame can be seen in the context of the so-called superselection rules \cite{ReviewQuantumRef}. While a selection rule prohibits the coupling of two eingenstates of a Hamiltonian, that is $\bra{\psi_1}H\ket{\psi_2} = 0$, which, consequently, makes it impossible to overlap these eigenstates, a superselection rule, in turn, is an extension of this concept to any other observable, $\bra{\psi_1}A\ket{\psi_2} = 0$, thus prohibiting the preparation of a superposition of these states in any circumstance \cite{giuilini2009}. These rules were originally introduced as fundamental in the context of non-relativistic quantum mechanics. The typical example is the superselection rule with respect to electrical charge that prohibits the preparation of quantum states with superposition of charge eigenstates \cite{wickwightmanwigner1970}. However, as Aharanov and Susskind argued \cite{aharonovsusskind1967}, the axiomatic introduction of these rules is based on the non-physical principle which says that there are absolute operations without any dependence on a reference system. Once a reference frame is introduced, from a practical point of view the problem can be circumvented and the preparation of overlapping states becomes possible again. However, the possibility of fundamental superselection rules remains still open. For our purposes we will focus on the practical approach of these rules as restrictions that arise due to the absence of an appropriate framework for the description of the system. 

In this context, the Page-Wootters mechanism that was initially proposed as an explanation to the emergence of time in an universe where a fundamental superselection rule for energy exists \cite{pagewootters1983,wootters1984,Pegg_1991} can be viewed as a quantum clock mechanism in a quantum reference frame perspective \cite{tiagomartinelli2019}. Adopting this perspective, in this paper we seek to understand the resource behind the operation of the Page-Wootters clock (PWC) by using a particular proposal to quantify how well a quantum system can stand as a reference frame for other quantum system, known as \textit{shared asymmetry} \cite{vaccaro2008}. Following the same path of Ref. \cite{tiagomartinelli2019} we show that besides the necessary existence of the \textit{mutual asymmetry}, which in this case turns to be \textit{internal coherence} \cite{kwonjeongjennings2018, leandro2018}, there is a kind of \textit{internal entanglement} that plays a role in the operation of PWC and in any quantum reference frame described by the $U(1)$ compact group.     

The paper is organized as follows. In Section II we introduce the Page-Wootters model of quantum clock from the perspective of quantum reference frames and exemplify it with a qubit clock. Section III is devoted to introducing the informational measure of shared asymmetry applied to PWC and some previous results and examples. In Section IV we present some extensions of the previous examples, discussing what they have in common, and finally in Section V, based on the previous examples and discussions, we present the main result of this work: \textit{the relation between the shared asymmetry for the $U(1)$ compact group and the relative entropy of entanglement}. We discuss some implications of this result and in Section VI we present our conclusions and suggestions for possible further studies.

\section{PWC and Quantum Reference Frames}

In 1983, Don Page and William Wootters proposed that given the possibility of the existence of a fundamental superselection rule for energy, in the sense that the whole universe could be described as a closed and stationary system, the reason why we observe temporal evolution in the world would be due to correlations between subsystems where one would act as a time reference frame for the other in a relational way \cite{pagewootters1983}. This perspective consists of an internalization of the reference frame in the system from an extension of Hilbert's space from $\mathcal{H}_S$ to $\mathcal{H}_S\otimes\mathcal{H}_R$, where $\mathcal{H}_R$ is space of the reference frame. In this case it is a clock which, in principle, can be seen as a type of phase reference frame described by the action of the group $U(1)$ on the space by unitary representation. 	

Since time is considered an inaccessible parameter in the Page-Wootters model, we have a global symmetry in relation to it so that the total Hamiltonian $H = H_S\otimes\mathds{1}_R + \mathds{1}_S\otimes H_R$ satisfies
\begin{equation}
	H\ket{\psi}_{SR} = 0,
	\label{eq_wheeler-witt}
\end{equation}
where $\ket{\psi}\in\mathcal{H}_S\otimes\mathcal{H}_R$ and $\mathds{1}_{\alpha}$ the identity operator in the subsystems $\alpha = S,R$. The same equation can be written in the density operator formalism. From Eq. (\ref{eq_wheeler-witt}), we have that, for $U_t = e^{-iHt}$, $U_t\rho_{SR} U_t^{\dagger} = \rho_{SR}$, which implies that $[U_t, \rho_{SR}]=0$. Being $e^{-iHt} = \sum\limits_{k=0}^\infty\frac{(-it)^k}{k!}H^k$ due to the commutation of $U_t$ with $\rho_{SR}$ it follows that
\begin{equation}
	[H,\rho_{SR}] = 0,
\end{equation}
where $\rho_{SR}\in\mathcal{B}(\mathcal{H}_S\otimes\mathcal{H}_R)$ (the space of bounded operators on $\mathcal{H}_S\otimes\mathcal{H}_R$).

The absence of an adequate framework obliges us to describe our states as an average with respect to all possible parameters, thus making the resulting system symmetrical and, consequently, frameworkless. In this case, considering the situation where both the system and the reference have equidistant spectrum, i.e., their energies are integer multiples of some constant value, this average will be a time average which, being also described by the group $U(1)$, can be written as
\begin{equation}
	\mathcal{G}(\rho_{SR}) = \int_0^{2\pi}\frac{d\phi}{2\pi}\,\mathcal{U}^{S,R}_{\phi}(\rho_{SR}), 
	\label{G-twirling_page_wootters}
\end{equation} 
where $\mathcal{U}^{S,R}_{\phi}(\bullet)=U^S_{\phi}\otimes U^R_{\phi}(\bullet) U^{R\dagger}_{\phi}\otimes U^{S\dagger}_{\phi}$, with $U_{\phi}^{\alpha} = \{e^{-i\phi H_{\alpha}};\phi\in[0,2\pi]\}$, $\alpha=S,R$, and $\frac{d\phi}{2\pi}$ is the Haar measure \cite{barutgroup}. As a consequence, we physically internalize the referential as part of the total system $S+R$, which allows us to describe the system $S$ in terms of the states of the time reference system $R$ that will serve as the ``hands of the clock''. To build these states, we start from an initial state $\ket{\psi_R(0)}$ associated with the identity element of the $U(1)$ group, which in this case is the zero element. From the representation $U^R_{\phi} = e^{-iH_R\phi}$ we can generate the remaining states by applying $U^R_{\phi}$ to the initial state, since $U^R_{\phi}\ket{\psi_R(\phi^{'})} = \ket{\psi_R(\phi + \phi^{'})}$, $\forall\, \phi, \phi^{'}\in U(1)$. The only restriction is that the base generated in this way is orthonormal, that is, $\braket{\psi_R(\phi^{'})|\psi_R(\phi + \phi^{'})} = \delta(\phi - (\phi+\phi'))$, because it allows us to distinguish the states perfectly \cite{changingqrf}. 

Consider a simple illustrative model, based on qubits, given by the following Hamiltonian
\begin{equation}
	H = \sigma_z^R\otimes \mathds{1}^S + \mathds{1}^R\otimes\sigma_z^S,
\end{equation}
where $\sigma_z^{\alpha}$ is the Pauli operator in the $\hat{z}$ direction. In this model the state $\ket{+} = (\ket{0} + \ket{1})/\sqrt{2}$ is adopted as symbolizing the ``12 o'clock'' and the state $\ket{-} = (\ket{0} - \ket{1})/\sqrt{2}$ as symbolizing the ``6 o'clock'', therefore separated by a phase $\pi$, which guarantees us distinguishability. Thus, the evolution of states is given by
\begin{equation}
	e^{-i\frac{\pi}{2}\sigma_z}\ket{+} = \ket{-}
\end{equation}
and
\begin{equation}
	e^{-i\frac{\pi}{2}\sigma_z}\ket{-} = \ket{+},
\end{equation} 
disregarding the global phase. By associating this reference system $R$ to a system $S$ also given by a qubit on the same basis, 
the dynamics of $S$ starts to be described through $R$ according to the associations we made between the states of $R$ and the markings of a clock. This consequence that allow us to recover the dynamics of the Schr\"{o}dinger equation for the $S$ system by the alignment of the $S$ and $R$ systems through the formality of conditional probabilities \cite{giovannettilloydmaccone2015}. 

In general, however, this model presents a certain uncertainty in the $R$ markings, in the sense that for a given state of $S$ we can have non-zero probabilities both in relation to $\ket{+}$ and $\ket{-}$, which does not allow us to associate precisely 12 or 6 hours with the status of $S$. This uncertainty will vary depending on the composite system used, and may even reach zero for some specific cases. The question then arises: How can we measure if the system $R$ serves as a benchmark for the system $S$? This question can be answered by information measure known as \textit{shared asymmetry}.

\section{Shared asymmetry and Page-Wootters}
\label{sec_sh_asymmetry}

Proposed in Ref. \cite{vaccaro2008}, shared asymmetry is a quantifier of how asymmetric one subsystem is in relation to the other in a bipartite state, or putting it another way a quantifier of the correlations between the asymmetries for each part. This measure is defined by,
\begin{equation}
	\mathcal{A}_{G\otimes G}^{(sh)}(\rho_{SR}) = \mathcal{S}(\mathcal{G}_{G\otimes G}(\rho_{SR})) - \mathcal{S}(\mathcal{G}(\rho_{SR})).
	\label{shared_asymmetry}
\end{equation}
where $\mathcal{S}(\rho) = - \Tr(\rho\log\rho) = -\sum\limits_x\lambda_x\log\lambda_x$, being log with base $2$, is the von Neumann entropy and 
\begin{equation}
	\mathcal{G}_{G\otimes G}(\rho_{SR}) = \int_0^{2\pi}\frac{d\phi}{2\pi}\int_0^{2\pi}\frac{d\phi'}{2\pi}\,\mathcal{U}^{S,R}_{\phi,\phi'}(\rho_{SR}), 
	\label{G-twirling_local_page_wootters}
\end{equation} 
is a local average (while Eq. (\ref{G-twirling_page_wootters}) is a global average) for the $U(1)$ group, being $\mathcal{U}^{S,R}_{\phi,\phi'}(\cdot)=U^S_{\phi}\otimes U^R_{\phi'}(\cdot) U^{R\dagger}_{\phi}\otimes U^{S\dagger}_{\phi'}$.
The resource quantified by this measure is known in the literature as \textit{shared reference frame} \cite{ReviewQuantumRef} and it is used when two parts in different laboratories do not have the same reference frame and, therefore, are under a local superselection rule, because the parts are limited to the preparation of local states. An interesting example of this constraint is a state of the type $\ket{+}_A\ket{+}_B$, known as \textit{refbit} \cite{refbit2005} that even being a product state in the absence of a common reference between Alice and Bob cannot be produced.

However, this measure can also allow us to evaluate how well a subsystem serves as a reference for another subsystem. Recently, in Ref. \cite{tiagomartinelli2019}, this potential was explored  
in the case of product states to quantify, within the context of the Page-Wootters mechanism, how well a subsystem $R$ serves as a clock system for a subsystem $S$. An important result of this application was the definition of upper bounds for shared asymmetry. In this way, it is possible to know for what value we will have the best possible case of internal reference frame, which means to say the greatest possible asymmetry between the subsystems $S$ and $R$. However, before we can enunciate and prove this result it is necessary to show some manipulations that can be done in the equation of shared asymmetry, which will be useful later.
 
First, for a product state $\rho_{SR}=\rho_S\otimes\rho_R$, we can rewrite the shared asymmetry equation as follows
\begin{eqnarray}
	\mathcal{A}_{G\otimes G}^{(sh)}(\rho_{SR}) &=& \mathcal{S}(\mathcal{G}_{G\otimes G}(\rho_{SR})) - \mathcal{S}(\mathcal{G}(\rho_{SR})) \nonumber \\
	&=& \mathcal{S}(\mathcal{G}(\rho_S)\otimes\mathcal{G}(\rho_R)) - \mathcal{S}(\mathcal{G}(\rho_{SR})) \nonumber \\
	&=& [\mathcal{S}(\mathcal{G}(\rho_S)) + \mathcal{S}(\mathcal{G}(\rho_R)) - \mathcal{S}(\mathcal{G}(\rho_{SR}))] \nonumber \\
	&-& [\mathcal{S}(\rho_S) + \mathcal{S}(\rho_R) -\mathcal{S}(\rho_{SR})] \nonumber \\
	&=& A_G(\rho_S) + A_G(\rho_R) - A_G(\rho_{SR}),
	\label{ass_mutua}
\end{eqnarray}  
where $A_G(\cdot)$ is the Holevo asymmetry \cite{marvianspekkens2016} given by
\begin{equation}
	A_G(\rho) = \mathcal{S}(\mathcal{G}(\rho)) - \mathcal{S}(\rho).
\end{equation}
There is another way of writing this equation, for a more general state $\rho_{SR}$, not necessary a product state or a pure state. For this, we use that the unitary action of any compact Lie group in the Hilbert space allows one to break it down into a direct sum of charge sectors which carry irreducible representations of the group \cite{ReviewQuantumRef}: $\mathcal{H}_S\otimes\mathcal{H}_R=\bigoplus_M\mathcal{H}_M$. In the case of $U(1)$ group this sectors do not need to be again broken down in the gauge and multiplicity virtual subsystems, $\mathcal{M}_M\otimes\mathcal{N}_M$, because the irreducible representations of this group are one-dimensional and so the gauge subsystem $\mathcal{M}_M$ will be trivial and can be disregarded. Thus, we can decompose the Hilbert spaces of the system $S$ and the clock  $R$ as $\mathcal{H}_S = \bigoplus_{m_S}\mathcal{H}_{m_S}$ and $\mathcal{H}_R = \bigoplus_{m_R}\mathcal{H}_{m_R}$, respectively, thus $\mathcal{H}_S\otimes\mathcal{H}_R=\bigoplus_{M=m_S+m_R}\mathcal{H}_M$. Being $\Pi_{m_S}$ and $\Pi_{m_R}$ the projectors onto the charge sectors $\mathcal{H}_{m_S}$ and $\mathcal{H}_{m_R}$, we have $\Pi_M=\sum_{M=m_S+m_R}\Pi_{m_S}\otimes\Pi_{m_R}$ as a projector onto the the charge sectors $\mathcal{H}_M$ of the total Hilbert $\mathcal{H}_S\otimes\mathcal{H}_R$. Using them, we can rewrite the global and the local average (see the Appendix \ref{Apendice_A} for more details) as
\begin{eqnarray}
	\mathcal{G}(\rho_{SR}) &=& \sum_{M}\Pi_M\rho_{SR}\Pi_M \nonumber \\
						   &=& \Pi_{G}(\rho_{SR})
						   \label{dephasing}
\end{eqnarray} 
where $\Pi_{G}(\cdot)$ is known as \textit{dephasing operator} and
\begin{eqnarray}
	\mathcal{G}_{G\otimes G}(\rho_{SR}) &=& \sum_{m_S,m_R}(\Pi_{m_S}\otimes\Pi_{m_R})\rho_{SR}(\Pi_{m_S}\otimes\Pi_{m_R}) \nonumber \\
	                                    &=& \Delta(\rho_{SR})
	                                    \label{fully_dephasing}, 
\end{eqnarray}
where $\Delta(\cdot)$ is known as \textit{fully dephasing operator}.
As was mentioned in Ref. \cite{marvianspekkens2016} the resource theory of asymmetry turns to be equivalent to resource theory of coherence when the asymmetry is defined with respect to an Abelian group and thus simply correspond to coherence between some preferred set of subspaces. In our case, as we are dealing with time-translations asymmetry, the charge sectors are the eigenspaces of the Hamiltonian, so $\Pi_{G}(\cdot)$ and $\Delta(\cdot)$ are the maps that eliminate the \textit{external} coherence, i.e., the coherence outside the charge sectors $\mathcal{H}_M$, or in this case, the coherence between states of different energies, and the total coherence of the state, respectively \cite{kwonjeong2018}. Replacing Eqs. (\ref{dephasing}) and (\ref{fully_dephasing}) in the shared asymmetry, Eq. (\ref{shared_asymmetry}), we obtain an internal (or mutual) coherence quantifier $\mathcal{C}(S:R)$ \cite{leandro2018}
\begin{equation}
	\mathcal{A}_{G\otimes G}(\rho_{SR}) = \mathcal{C}(S:R) = S(\Delta(\rho_{SR})) - S(\Pi_G(\rho_{SR})).
\end{equation} 
The state after the dephasing operation is always a block-diagonal state in the energy basis, thus the measure $\mathcal{C}(S:R)$ permits to quantify the coherences that appears inside each block, i.e., coherence between quantum states of the same energy $M=m_s+m_R$.   

Now, using this results, especially Eqs. (\ref{ass_mutua}, (\ref{dephasing}) and (\ref{fully_dephasing}), we 
can enunciate and demonstrate the following lemma according to Ref. \cite{tiagomartinelli2019}.

\begin{lema}
	\label{eq_lema_upper}
	Shared asymmetry for a pure state $\rho_{SR}=\rho_S\otimes\rho_R$ satisfies the following bounds:
	\begin{equation}
		0\leq \mathcal{A}_{G\otimes G}^{(sh)}(\rho_{SR})\leq \min\{A_G(\rho_S), A_G(\rho_R)\}.
	\end{equation}
\end{lema}
\textit{Proof}: the first inequality is a direct consequence of the definition of the measure and  for the second we will use the property that Holevo monotones \cite{marvianspekkens2016}, $A_G$, do not increase under the action of the partial trace \cite{SchumacherWestmorelandWootters}, therefore
\begin{equation}
	A_G(\rho_S\otimes\rho_R) \geq A_G(\rho_{\alpha}), \, \alpha=S,R.
\end{equation} 
The equality can be shown by the following: choose a normalized state in $R$ on a eigenspace of a sufficiently large size compared to $S$ , that is, for $\rho_R\propto\Pi_{m_R}$, with $m_R\approx M$, being $\mathcal{G}(\rho_S\otimes\rho_R)=\sum_M\Pi_M(\rho_S\otimes\rho_R)\Pi_M$, $\Pi_M = \sum_{m_S+m_R=M}\Pi^S_{m_S}\otimes\Pi^R_{m_R}$, we will have
\begin{equation}
	A_G(\rho_S\otimes\rho_R)\approx\mathcal{S}(\rho_S) + \mathcal{S}(\mathcal{G}(\rho_R)) - \mathcal{S}(\rho_S)-\mathcal{S}(\rho_R) = A_G(\rho_R),
\end{equation} 
in which, due to the condition established with respect to the $R$ dimension, we use that $\mathcal{G}(\rho_S\otimes\rho_R)\propto\Pi_{m_S}\rho_S\Pi_{m_S}\otimes\sum_{m_R}\Pi_{m_R}\rho_R\Pi_{m_R}$ and the fact that entropy is additive.
Putting all these considerations together and using Eq. (\ref{ass_mutua}) we get
\begin{equation}
	\mathcal{A}_{G\otimes G}^{(sh)}(\rho_{SR}) = A_G(\rho_S) + A_G(\rho_R) - A_G(\rho_R) = A_G(\rho_S), 
\end{equation}
finishing the proof.

Based on the definition of this upper bound, Ref. \cite{tiagomartinelli2019} presents three examples of $S+R$ systems and calculates the shared asymmetry of these systems to compare with what would be the upper bound expected by the lemma \ref{eq_lema_upper}. We will present these examples in the following because our objective is to extend them to more general cases, which will lead us to a deeper analysis of the shared asymmetry for the group $U(1)$. 	

\begin{itemize}
	\item[(i)] Qubit model: Let both system $S$ and clock $R$ be given by a qubit $\rho_{\alpha} = \ket{+}\bra{+}$, where $\ket{+} = (\ket{0} + \ket{1}/\sqrt{2})$, being asymmetric in relation to the unitary representation $U_{\phi}^{\alpha} = \{e^{i\phi\sigma_z^{\alpha}};\phi\in[0, 2\pi]\}$, $\alpha=S,R$. An external observer under the action of a global superselection rule whose symmetry is represented by $U_{\phi} = \{e^{i\phi\sigma_z};\phi\in[0, 2\pi]\}$, with $\sigma_z = \sigma_z^S\otimes\mathds{1}_R + \mathds{1}_S\otimes\sigma_z^R$, will have access only to the degrees of freedom of the total system $\rho_S\otimes\rho_R$ independent of reference frames, that is
	\begin{equation}
		\mathcal{G}(\rho_S\otimes\rho_R) = \frac{1}{2\pi}\int_0^{2\pi} \, d\phi \, U_{\phi}(\rho_S\otimes\rho_R)U_{\phi}^{\dagger}. 	
	\end{equation}	  
The result, written in matrix form in the base $\{\ket{00}, \ket{01}, \ket{10}, \ket{11}\}$, will have the block-diagonal form
\begin{equation}
	\mathcal{G}[\rho_S\otimes\rho_R] = \frac{1}{4}\left[\begin{array}{cccc}
    1 & & &\\
    &1 &1 &\\
    &1 &1 &\\
    & & &1\\
\end{array}\right],\\
	\label{matriz_pagewootters}
\end{equation}
where all the other entries that do not appear are considered to be null. The shared asymmetry of this state, using von Neumann entropy given by $\mathcal{S}(\rho) = -\sum_x \lambda_x\log\lambda_x$, being $\lambda_x$ the eigenvalues of $\rho$, will be
\begin{eqnarray}
	\mathcal{A}_{G\otimes G}^{(sh)}(\rho_S\otimes\rho_R) &=& \frac{1}{2},
\end{eqnarray}
Comparing with the maximum value for this system based on the lemma \ref{eq_lema_upper}, $A_G(\rho_S) = A_G(\rho_R) = 1$, we see that $\rho_S\otimes\rho_R$ does not saturate this bound, however shows the functioning of the Page-Wootters mechanism to describe time in a universe of two qubits, as proposed in Ref. \cite{wootters1984}

\item[(ii)]High reference location: it is expected that the larger the dimension of the reference frame $R$, the better it will guide the system $S$ \cite{ReviewQuantumRef}. We can see this considering that the system $S$ is given by the qubit $\ket{\psi_S} = (\ket{0} + \ket{1})/\sqrt{2})$ while the clock system $R$ is given by a qudit in a uniform superposition, also known as maximally coherent state,
\begin{equation}
	\ket{\psi_R} = \frac{1}{\sqrt{d}}\sum\limits_{m=0}^{d-1}\ket{m},
	\label{eq_max_verossiilhanca}
\end{equation}
with Hamiltonian $H_R = \sum_{m=0}^{d-1}m\ket{m}\bra{m}$. Writing the result of $\mathcal{G}(\rho_S\otimes\rho_R)$ in the matrix form using the basis $\{\ket{00}, \ket{01},...,\ket{0\,d-1}, \ket{10}, \ket{11},...,\ket{1\,d-1}\}$,
so that we always have together the labels that, added up, give the same total (for example, $\ket{01}$ and $\ket{10}$ are in sequence for both adding one), we will have
\begin{equation}
	\mathcal{G}(\rho_S\otimes\rho_R) = \frac{1}{2d}\left[\begin{array}{ccccccc}
    1 & & & & & &\\
    &1 &1 & & & &\\
    &1 &1 & & & &\\
    & & &\ddots & & & \\
    & & & &1 &1 & \\
    & & & &1 &1 &\\
    & & & & & &1 \\
\end{array}\right].\\
\label{eq_mat_2d}
\end{equation}
The shared asymmetry in this case will be
\begin{eqnarray}
	\mathcal{A}_{G\otimes G}^{(sh)}(\rho_S\otimes\rho_R) &=& 1 - \frac{1}{d},
\end{eqnarray}
that is, for $d\rightarrow\infty$ we have $\mathcal{A}_{G\otimes G}^{(sh)}(\rho_S\otimes\rho_R)\rightarrow 1 = A_G(\rho_S)$. Therefore, for this case, considering the size of $R$ large, we have a system that tends to saturate the upper bound and, consequently, is the best possible.

	\item[(iii)] High order of coherence: maintaining the clock's state as the uniform superposition of Eq. (\ref{eq_max_verossiilhanca}), we will use as an asymmetric state a state of the type
	\begin{equation}
		\ket{\psi_S} = \frac{1}{\sqrt{2}}(\ket{0} + \ket{d-1}),
		\label{eq_est_hiato}
	\end{equation}
that is, with order of coherence $d-1$, where for order of coherence $k$ of a state $\rho = \sum_{m,n}\rho_{m,n}\ket{m}\bra{n}$ we understand 1-norm of the sum of the elements outside the diagonal with  $k=m-n$. States like Eq. (\ref{eq_est_hiato}) are also known as states with a gap in their spectrum.  Globally symmetrizing the composite system $\rho_S\otimes\rho_R$ according to the point of view of an external observer, we will have
	\begin{equation}
	\mathcal{G}(\rho_S\otimes\rho_R) = \frac{1}{2d}\left[\begin{array}{ccccccc}
    1 & & & & &\\
    &\ddots & & & &\\
    & &1 &1 & &\\
    & &1 &1 & & \\
    & & & &\ddots & \\
    & & & & &1 \\
	\end{array}\right].\\
	\label{eq_mat_2d_hiato}
	\end{equation}
The shared asymmetry of this state is
\begin{eqnarray}
	\mathcal{A}_{G\otimes G}^{(sh)}(\rho_S\otimes\rho_R) &=& \frac{1}{d}.
\end{eqnarray}	
We see that the situation here is opposite to the previous one: if we do $d\rightarrow\infty$ we will have $\mathcal{A}_{G\otimes G}^{(sh)}(\rho_S\otimes\rho_R)\rightarrow 0$, which is the lower bound. Therefore, considering the size of $R$ large, will tend to be the worst case possible, that is even using what would be the ideal clock system, we have that the Page-Wootters mechanism will not work.
\end{itemize}

\section{Shared asymmetry for more general cases}
\label{sec_general_ex}

In the two cases presented here we make use of the symmetries that appear in the block-diagonal structure of the density matrices of the states $\rho_{SR}$ after the application of the global average operation  $\mathcal{G}$ for $G=U(1)$. As we will see, this symmetries make the calculation of the shared asymmetry much easier and help us to better visualize the implications of varying the dimension of the systems or the interval between their eigenstates. It is worth mentioning that although we work with examples having uniform superposition, the block-diagonal structures of the density matrices will remain the same for non-uniform cases, will vary only the coefficients that constitute the blocks. 


\subsection{Maximally coherent states}
\label{subsec:geral_sem}
To analyse the case where the system $S$ and the reference frame $R$ are both described for maximally coherent states, with dimensions $d_S$ and $d_R$, respectively, where $d_S\leq d_R$, we will construct the pattern of $\mathcal{G}(\rho_{SR})$ in descending order of total dimension maintaining the dimension of $R$. Beginning with the case where the dimensions of $S$ and $R$ are equal to $d$, we will decrease the dimension of $S$ to observe the changes in the density matrix and thus identify the general pattern that will allow us to calculate the shared asymmetry. 

So, let the system $S$ given by $\ket{\psi_S}=\frac{1}{\sqrt{d}}\sum_{m_S=0}^{d-1} \ket{m_S}$ and the reference frame $R$ for $\ket{\psi_R}=\frac{1}{\sqrt{d}}\sum_{m_R=0}^{d-1} \ket{m_R}$ whose Hamiltonians are given respectively by $H_S = \sum_{m_S=0}^{d-1}m_S\ket{m_S}\bra{m_S}$ and $H_R = \sum_{m_R=0}^{d-1}m_R\ket{m_R}\bra{m_R}$. Applying $\mathcal{G}$ operation in this state and writing the result as a density matrix we have
\begin{equation}
	\mathcal{G}(\rho_{SR}) = \frac{1}{d^2}\left[\begin{array}{ccccccc}
    A_1 & & & & & &\\
    &A_2 & & & & &\\
    & &\ddots & & & &\\
    & & &A_d & & & \\
    & & & &\ddots & & \\
    & & & & &A_2 &\\
    & & & & & &A_1 \\
\end{array}\right],\\
\end{equation}
where $A_n$ is a $n\times n$ square matrix that has all entries equal to one. 

For $\ket{\psi_S}=\frac{1}{\sqrt{d-1}}\sum_{m_S=0}^{d-2} \ket{m_S}$ and $\ket{\psi_R}=\frac{1}{\sqrt{d}}\sum_{m_R=0}^{d-1} \ket{m_R}$, using the same notation as before, the calculation of the $\mathcal{G}(\rho_{SR})$ gives 
\begin{widetext}
\begin{equation}
    \mathcal{G}(\rho_{SR}) = \frac{1}{d(d-1)}\left[\begin{array}{cccccccc}
    A_1 & & & & & & &\\
    &A_2 & & & & & &\\
    & &\ddots & & & & &\\
    & & &A_{d-1} & & & &\\
    & & & &A_{d-1} & & &\\
    & & & & &\ddots & &\\
    & & & & & &A_1 &\\
    & & & & & & &A_2\\
    \end{array}\right].
    \label{eq_exemplo_lema}
\end{equation}
\end{widetext}
For $\ket{\psi_S}=\frac{1}{\sqrt{d-2}}\sum_{m_S=0}^{d-3} \ket{m_S}$ and $\ket{\psi_R}=\frac{1}{\sqrt{d}}\sum_{m_R=0}^{d-1} \ket{m_R}$, we have
\begin{widetext}
\begin{equation}
    \mathcal{G}(\rho_{SR}) = \frac{1}{d(d-2)}\left[\begin{array}{ccccccccc}
    A_1 & & & & & & & &\\
    &A_2 & & & & & & &\\
    & &\ddots & & & & & &\\
    & & &A_{d-2} & & & & &\\
    & & & &A_{d-2} & & & &\\
    & & & & &A_{d-2} & & &\\
    & & & & & &\ddots & &\\
    & & & & & & &A_2 &\\
    & & & & & & & &A_1\\    
    \end{array}\right].
\end{equation}
\end{widetext}
We can see that a pattern already seems to emerge. Lets also look at the other side of the spectrum, that is, from the smallest to the largest dimensions of the system $S$.  Let $\ket{\psi_S}=\frac{1}{\sqrt{2}}(\ket{0} + \ket{1})$ and $\ket{\psi_R}=\frac{1}{\sqrt{d}}\sum_{m_R=0}^{d-1} \ket{m_R}$, we have
\begin{equation}
    \mathcal{G}(\rho_{SR}) = \frac{1}{2d}\left[\begin{array}{ccccc}
    A_1 & & & &\\
    &A_2 & & &\\
    & &\ddots & &\\
    & & &A_2 &\\
    & & & &A_1\\
    \end{array}\right],
\end{equation}
with $d-1$ matrices $A_2$.    
For $\ket{\psi_S}=\frac{1}{\sqrt{3}}(\ket{0} + \ket{1} + \ket{2})$ and $\ket{\psi_R}=\frac{1}{\sqrt{d}}\sum_{m_R=0}^{d-1} \ket{m_R}$
\begin{equation}
	\mathcal{G}(\rho_{SR}) = \frac{1}{3d}\left[\begin{array}{ccccccc}
    A_1 & & & & &\\
    &A_2 & & & & &\\
    & &A_3 & & & &\\
    & & &\ddots & & &\\
    & & & &A_3 & &\\
    & & & & &A_2 &\\
    & & & & & &A_1\\
    \end{array}\right],
\end{equation}	
with $d-2$ matrices $A_3$.
Based in all the examples shown so far the following general pattern can be observed by inductive reasoning in the matrices $A_n$ which form the blocks of the density matrix of the system $\mathcal{G}(\rho_{SR})$ and its dimension  
\begin{itemize}
     \item If $\mbox{dim}(\rho_{SR}) = d$ x $2$, so we have $d-1$ matrices $A_2$.\\
     \item If $\mbox{dim}(\rho_{SR}) = d$ x $3$, so we have $d-2$ matrices $A_3$.
      \newline\vdots  
     \item If $\mbox{dim}(\rho_{SR}) = d$ x $(d-2)$, so we have $3$ matrices $A_{d-2}$.\\
     \item If $\mbox{dim}(\rho_{SR}) = d$ x $(d-1)$, so we have $2$ matrices $A_{d-1}$.\\
     \item If $\mbox{dim}(\rho_{SR}) = d$ x $d$, so we have one matrix $A_d$.
\end{itemize}
All the other matrices in each case will appear in pairs. 

Using all this information listed above and the property that matrices of this type have their eigenvalues always equal to $n$ with multiplicity $1$ and equal to $0$ with multiplicity $n-1$ we can write a general form to the von Neumann entropy of $\mathcal{G}(\rho_{SR})$ 
\begin{widetext}
 \begin{equation}
    S(\mathcal{G}(\rho_{SR})) = -\frac{1}{d_S d_R}\bigg[2\sum_{X_S=1}^{d_S-1} x_S \log \left(\frac{x_S}{d_R d_S}\right) + (d_R - d_S + 1)d_S \log \left(\frac{1}{d_R}\right)\bigg]
    \label{eq_vonneumann_twirl2},
 \end{equation} 
\end{widetext}    
    where $d_S$ is the dimension of $\rho_S$ and $d_R$ the dimension of $\rho_R$.

Replacing Eq. (\ref{eq_vonneumann_twirl2}) in the shared asymmetry equation, Eq. (\ref{shared_asymmetry}), and knowing that $S(\mathcal{G}_{G \otimes G}(\rho_{SR})) = S(\Delta(\rho_{SR})) = \log{d_S d_R}$, we will arrive at the following general equation for the shared asymmetry of maximally coherent systems with different dimensions
\begin{widetext}
\begin{equation}
      \mathcal{A}_{G\otimes G}^{(sh)}(\rho_{SR}) = \frac{2}{d_S d_R}\sum_{x_S=1}^{d_S-1}x_S\log x_S - \frac{d_S-1}{d_R}\log d_S + \log d_S.
      \label{eq_caso_dimdif}
\end{equation}
\end{widetext}

It is worth mentioning that this result is valid for systems without gaps. The upper bound in this case will be $\log d_S$ where $d_S \leq d_R$, since $\mathcal{A}_G(\rho_S) = \log d_S$ and $\mathcal{A}_G(\rho_R) = \log d_R$. For $d_R \xrightarrow{} \infty$, it is easy to see that $\mathcal{A}_{G\otimes G}^{(sh)} = \log d_S$. Therefore, for maximally coherent states, if the dimension of the reference frame $R$ tend to infinity, the upper bound of the shared asymmetry will be reached. Such a result, as previously stated, was already expected since some results in the literature pointed out that the larger the dimension of the reference frame system $R$, more the orientation of the system $S$ in relation to it is optimized. \cite{ReviewQuantumRef}  

\subsection{Maximally coherent state \textit{versus} qubits with gap}
\label{subsec:qdit_qdit_intervalo}
This is the most interesting case. As we described previously, it was shown in Ref. \cite{tiagomartinelli2019} that for a system $\rho_{SR}$ where the reference frame $R$ is in a maximally coherent state, $\ket{\psi_R}=\frac{1}{\sqrt{d}}\sum_{m_R=0}^{d-1} \ket{m_R}$, and the system $S$ is in an state of high coherence order of the type $\ket{\psi_R}=\frac{1}{\sqrt{2}}(\ket{0} + \ket{d-1})$, we have that if the dimension of $R$ goes to infinity the shared asymmetry of $\rho_{SR}$ will goes to zero, constituting in the worst possible scenario. On the other hand, if $S$ is a qubit like $\ket{\psi_S} = \frac{\ket{0}+\ket{1}}{\sqrt{2}}$ and $R$ is still the same, as we saw, as the dimension of $R$ goes to infinity, the measure goes to the maximum value, being the best possible scenario. Our aim here is to connect the shared asymmetry of the worst and the best case by varying the gap dimension in the system $S$ and then write it in a general way considering $S$ with any $d_g$ dimension gap. 

Let $\ket{\psi_R}=\frac{1}{\sqrt{d}}\sum_{m_R=0}^{d-1} \ket{m_R}$be the reference frame $R$ and  $\ket{\psi_S}=\frac{1}{\sqrt{2}}(\ket{0} + \ket{d-1})$ the system $S$ we will have that the density matrix of the state $\mathcal{G}(\rho_{SR})$ will be
\begin{equation}
        \mathcal{G}(\rho_{SR}) = \frac{1}{2d}\left[\begin{array}{cccccc}
        A_1 & & & &\\
        &\ddots & & &\\
        & &A_2 & &\\
        & & &\ddots & \\
        & & & &A_1 \\
        \end{array}\right].
\end{equation}
For a system $S$ with dimension gap a smaller unit, $\ket{\psi_S}=\frac{1}{\sqrt{2}}(\ket{0} + \ket{d-2})$, we will have 
\begin{equation}
        \mathcal{G}(\rho_{SR}) = \frac{1}{2d}\left[\begin{array}{ccccccc}
        A_1 & & & & &\\
        &\ddots & & & &\\
        & &A_2 & & &\\
        & & &A_2 & &\\
        & & & &\ddots &\\
        & & & & &A_1\\
        \end{array}\right].
\end{equation}
Again, here we can begin to find a pattern. So let us look at the opposite extreme, $\ket{\psi_S}=\frac{1}{\sqrt{2}}(\ket{0} + \ket{1})$, we will have (as already seen)
\begin{equation} 
		\mathcal{G}(\rho_{SR}) = \frac{1}{2d}\left[\begin{array}{ccccc}
        A_1 & & & &\\
        &A_2 & & &\\
        & &\ddots & &\\
        & & &A_2 &\\
        & & & &A_1\\
        \end{array}\right],
\end{equation}        
with $d-1$ matrices $A_2$. Finally for $\ket{\psi_S}=\frac{1}{\sqrt{2}}(\ket{0} + \ket{2})$
\begin{equation}
        \mathcal{G}(\rho_{SR}) = \frac{1}{2d}\left[\begin{array}{cccccccc}
        A_1 & & & & & &\\
        &A_1 & & & & &\\
        & &A_2 & & & &\\
        & & &\ddots & & &\\
        & & & &A_2 & &\\
        & & & & &A_1 &\\
        & & & & & &A_1\\
        \end{array}\right],
\end{equation}
with $d-2$ matrices $A_2$. We can see that the number of matrices $A_2$ clearly depends on the dimension of the gap in the qubit, ranging from $1$ to $d-1$ matrices for, respectively, dimension from $d-2$ to zero of the gap. With this it is possible to calculate $S(\mathcal{G}(\rho_{RS}))$ for the most general case of a gap with dimension $d_g$ by relating it to the number $n_m$ for matrices $A_2$. Let $\ket{\psi_R}=\frac{1}{\sqrt{d}}\sum_{m_R=0}^{d-1} \ket{m_R}$ and $\ket{\psi_S}=\frac{1}{\sqrt{2}}(\ket{0} + \ket{d-n_m})$, we will have
  \begin{equation}
      \mathcal{S}(\mathcal{G}(\rho_{RS})) = \log 2d - \frac{n_m}{d}
  \end{equation}
Consequently, knowing that $S(\mathcal{G}_{G\otimes G}(\rho_{SR})) = \log 2d$, the shared asymmetry, Eq. (\ref{shared_asymmetry}), will be
  \begin{equation}
      \mathcal{A}_{G\otimes G}^{(sh)}(\rho_{SR}) = \frac{n_m}{d}, 
  \end{equation}

Applying this equation to the extreme cases we recovery and connect the results of Ref. \cite{tiagomartinelli2019}. This allows us to visualize the clear dependence on shared asymmetry in relation to the block-diagonal structure of the density matrix of the state $\rho_{SR}$. Bearing this in mind, in the next section we will focus in investigate more deeply the structure of each block, trying to comprehend what constitutes the resource quantified by the shared asymmetry that turns a system into a reference frame for the other. Such blocks will be associated with states the we will call \textit{internal states} of the system $\rho_{SR}$.  

\section{Shared asymmetry: an analysis for charge sector of the Hilbert space}

Because we are dealing with the $U(1)$ group, each one of the blocks $A_n$ that appears in the density matrix of $\mathcal{G}(\rho_{SR})$ is related to a charge sector $\mathcal{H}_M$ of the total Hilbert space decomposition $\mathcal{H}_S\otimes\mathcal{H}_R=\bigoplus_{M=m_S+m_R}\mathcal{H}_M$. Thus, to analyse the role of the block-diagonal structure of these states in the shared asymmetry is interesting to do an analysis of each charge sector of the total Hilbert space. For that, we can use to the simplest formulation of the global and local average in terms of the dephasing and fully dephasing operations (Eqs. (\ref{dephasing}) and (\ref{fully_dephasing})), which, to recap, are $\mathcal{G}(\rho_{SR})=\sum_M\Pi_M\rho_{SR}\Pi_M$ where for global we have $\Pi_M=\sum_{M=m_S+m_R}\Pi_{m_S}\otimes\Pi_{m_R}$) and  for local $\Pi_M=\sum_{m_S,m_R}\Pi_{m_S}\otimes\Pi_{m_R}$. 
Replacing them in the shared asymmetry equation, we will have
\begin{widetext}
\begin{eqnarray}
	\mathcal{A}_{G\otimes G}^{(sh)}(\rho_{SR}) &=& \mathcal{S}\left(\sum_{m_S,m_R}(\Pi_{m_S}\otimes\Pi_{m_R})\rho_{SR}(\Pi_{m_S}\otimes\Pi_{m_R})\right) \nonumber \\
	 &-& \mathcal{S}\left(\sum_M\sum\limits_{\substack{(m_{S_1} + m_{R_1}), \\ (m_{S_2} + m_{R_2})} = M}\left(\Pi_{m_{S_1}}\otimes\Pi_{m_{R_1}}\right)\rho_{SR}\left(\Pi_{m_{S_2}}\otimes\Pi_{m_{R_2}}\right)\right)
\end{eqnarray}
\end{widetext}
As we want to study the most general case possible where both the system and the reference have equidistant spectrum, we will consider both the state of the system $S$ and that of the reference frame $R$ given by a Hamiltonian of the type $H = \sum_{m=0}^{d-1}m\ket{m}\bra{m}$, so that by expanding the global state $\rho_{SR}$ in the same basis of the Hamiltonian we will get
\begin{equation}
	\rho_{SR} = \sum_{\substack{m_{S_1}, m_{R_1}, \\ m_{S_2}, m_{R_2}} = 0}^{d-1}c_{m_{S_1}, m_{R_1}, m_{S_2}, m_{R_2}}\ket{m_{S_1}, m_{R_1}}\bra{m_{S_2}, m_{R_2}}.
\end{equation}  
Substituting this form of the state in the previous equation and applying the projections, being $c_{m_{S_1}, m_{R_1}} := c_{m_{S_1}, m_{R_1}, m_{S_2}, m_{R_2}}$, for $m_{S_1} = m_{S_2}$ and $m_{R_1} = m_{R_2}$, we will obtain
\begin{widetext}
\begin{eqnarray}
	\mathcal{A}_{G\otimes G}^{(sh)}(\rho_{SR}) &=&  \mathcal{S}\left(\sum_{m_{S_1},m_{R_1}}c_{m_{S_1}, m_{R_1}}\ket{m_{S_1}, m_{R_1}}\bra{m_{S_1}, m_{R_1}}\right) \nonumber \\ 
	&-& \mathcal{S}\left(\sum_M\sum\limits_{\substack{(m_{S_1} + m_{R_1}), \\ (m_{S_2} + m_{R_2})} = M}c_{m_{S_1}, m_{R_1}, m_{S_2}, m_{R_2}}\ket{m_{S_1}, m_{R_1}}\bra{m_{S_2}, m_{R_2}}\right),
	\label{eq_setor_carga}
\end{eqnarray}
\end{widetext}
We can see that while von Neumann entropy second argument is written in terms of the charge sector $\mathcal{H}_M$, the first is written in terms of the local charge sectors $\mathcal{H}_{m_S}$ and $\mathcal{H}_{m_R}$ from the Hilbert spaces $\mathcal{H}_S$ and $\mathcal{H}_R$, respectively. As all local symmetric states are also globally symmetric, at least when $G=U(1)$, that is $\mathcal{G}_G(\mathcal{G}_{G\otimes G}(\rho_{SR})) = \mathcal{G}_{G\otimes G}(\rho_{SR}) \rightarrow\Pi_G(\Delta(\rho_{SR}))=\Delta(\rho_{SR})$ (because the fully dephasing operation $\Delta$ transforms the system in a diagonal state whereas the global dephasing operation $\Pi_G$ transforms the system in a block-diagonal state, both on the same basis), we can project the argument of the first term of R.H.S of Eq. (\ref{eq_setor_carga}) in terms of the charge sectors $\mathcal{H_M}$. This will lead us to
\begin{widetext}
\begin{eqnarray}
	\label{eq_ass_shared2}
	\mathcal{A}_{G\otimes G}^{(sh)}(\rho_{SR}) &=& \mathcal{S}\left(\sum_M\sum_{m_{S_1} + m_{R_1} = M}c_{m_{S_1}, m_{R_1}}\ket{m_{S_1}, m_{R_1}}\bra{m_{S_1}, m_{R_1}}\right) \nonumber \\ 
	&-& \mathcal{S}\left(\sum_M\sum\limits_{\substack{(m_{S_1} + m_{R_1}), \\ (m_{S_2} + m_{R_2})} = M}c_{m_{S_1}, m_{R_1}, m_{S_2}, m_{R_2}}\ket{m_{S_1}, m_{R_1}}\bra{m_{S_2}, m_{R_2}}\right)
\end{eqnarray}
\end{widetext}
Using that $M = m_S + m_R$, we will make the following label change in the states above: $\ket{m_S, m_R}\rightarrow\ket{m_S, M - m_S}$, which consequently will lead us to also change the sum that couple the $m$ indices. With these changes, we denote 
\begin{equation}
	\rho_M^{un} \equiv \sum\limits_{m_{S_1}, m_{S_2} = 0}^M c_{m_{S_1}, m_{S_2}}\ket{m_{S_1}, M - m_{S_1}}\bra{m_{S_2}, M - m_{S_2}}
\end{equation}
and, being $c_{m_{S_1}}:=c_{m_{S_1}, m_{S_2}}$ for $m_{S_1}=m_{S_2}$,
\begin{equation}
	\rho_M^{un'} \equiv \sum_{m_{S_1} = 0}^M c_{m_{S_1}}\ket{m_{S_1}, M - m_{S_1}}\bra{m_{S_1}, M - m_{S_1}},
\end{equation}
where $\rho_M^{un}, \rho_M^{un'}\in\mathcal{B}(\mathcal{H}_M)$,  and these states are unnormalized. However, we can normalize these states multiplying both by $\frac{\Tr(\rho_M^{un})}{\Tr(\rho_M^{un})}$ because $\Tr(\rho_M^{un}) = \Tr(\rho_M^{un'})$. Doing so and substituting in Eq. (\ref{eq_ass_shared2}), we find
\begin{equation}
	\label{eq_ass_shared3}
	\mathcal{A}_{G\otimes G}^{(sh)}(\rho_{SR}) = \mathcal{S}\left(\sum_M\Tr(\rho_M^{un})\rho_M'\right) 
	- \mathcal{S}\left(\sum_M\Tr(\rho_M^{un})\rho_M\right)
\end{equation}
where $\rho_M' \equiv \frac{\rho_M^{un'}}{\Tr(\rho_M^{un})}$ and $\rho_M \equiv \frac{\rho_M^{un}}{\Tr(\rho_M^{un})}$. Now using the von Neumann entropy property that $\mathcal{S}\left(\sum_i p_i\rho_i\right) = H(p_i) + \sum_i p_i\mathcal{S}(\rho_i)$ when $\rho_i$ have orthogonal support \cite{nielsen2002quantum}, which is the case here, we can rewrite the above equation as follows 
\begin{eqnarray}
	\mathcal{A}_{G\otimes G}^{(sh)}(\rho_{SR}) &=& \sum_M\Tr(\rho_M^{un})\left(\mathcal{S}(\rho_M') - \mathcal{S}(\rho_M)\right) \\
	\label{eq_ass_sh_m}
	&=& \sum_M\Tr(\rho_M^{un})\left(\mathcal{A}_{G\otimes G}^{(sh)^M}(\rho_M)\right),
\end{eqnarray}
and from the first to the second equation we use that $\mathcal{G}_{G\otimes G}(\rho_M') = \rho_M' $ and $\mathcal{G}_G(\rho_M) = \rho_M$.

We got here to the point where we wanted to. By writing the shared asymmetry of the system $\rho_{SR}$ as a summation of shared asymmetries of its internal states $\rho_M$ on the charge sectors $\mathcal{H}_M$, we can enunciate the follow theorem
\begin{teo*}
	Let the shared asymmetry for $G=U(1)$ given by Eq. (\ref{eq_ass_sh_m}). We then have that
	  \begin{equation}
	  	\label{eq_teorema}
	  	\mathcal{A}_{G\otimes G}^{(sh)^M}(\rho_M) = E_{R}(\rho_M)
	  \end{equation}
	  where $E_{R}(\rho) = \min_{\sigma \in \text{SEP}}\mathcal{S}(\rho\parallel\sigma)$, being $\text{SEP}$ the set of all separable states, is the relative entropy of entanglement. Therefore,
	  \begin{equation}
	  	\label{eq_teorema2}
	  	\mathcal{A}^{(sh)}_{G\otimes G}(\rho_{SR}) = \sum_M\Tr(\rho_M^{un})(E_R(\rho_M))
	  \end{equation}	   
\end{teo*}  
Proof: A well-known result in the area of quantum information, especially in studies of entanglement quantifiers, is the so-called Vedral-Plenio Theorem \cite{vedralplenio1998}. Such theorem demonstrates that the relative entropy of entanglement for any pure bipartite state $\rho = \sum\limits_{n_1,n_2}\sqrt{c_{n_1}c_{n_2}}\ket{\phi_{n_1},\psi_{n_1}}\bra{\phi_{n_2},\psi_{n_2}}$ is equal to the von Neumann reduced entropy of the same state, given by $E(\rho) = -\sum_n c_n\ln c_n$. To make this demonstration, they proved that the closest separable state of $\rho$ that minimizes the relative entropy is precisely its diagonal version, that is $\rho^{'} = \sum\limits_n c_n\ket{\phi_n,\psi_n}\bra{\phi_n,\psi_n}$. Using some extended versions of this theorem (one of which is shown in the appendix B) it is possible to show that for some classes of mixed states, including the type 
\begin{equation}
	\rho = \sum\limits_{{n_1,n_2} = 0}^N c_{n_1,n_2}\ket{n_1;N-n_1}\bra{n_2;N-n_2},
	\label{eq_estados_B1}
\end{equation}
the closest separable state is the diagonal state $\rho^{'} = \sum\limits_{{n_1} = 0}^N c_{n_1}\ket{n_1;N-n_1}\bra{n_1;N-n_1}$. As this is the case with $\rho_M$ e $\rho_M'$ this proves Eq. (\ref{eq_teorema}) and consequently Eq. (\ref{eq_teorema2}).

This entanglement will be called \textit{internal entanglement}, since it consists of entanglement of the states $\rho_M\in\mathcal{B}(\mathcal{H}_M)$ which we called \textit{internal states} of the system $\rho_{SR}$. The curious thing about this result is that this kind of entanglement can appear even if the state $\rho_{SR}$ itself is separable. Actually, as mentioned in Ref. \cite{werner1989} the average operation $\mathcal{G}$ (global or local) applied in a classical correlated state can only generate classical correlated states. So, all the examples we have worked on in sections \ref{sec_general_ex} and \ref{sec_sh_asymmetry}, are in fact separable states and have entangled internal states $\rho_M$.  


From the theorem we can now revisit some of the previous results of the literature, interpreting then for the perspective of the concept of internal entanglement:
\begin{itemize}
	\item \textbf{Entanglement and the Page-Wootters mechanism:} Although the idea that the resource responsible for the operation of the Page-Wootters mechanism was entanglement had been propagated for a long time, recently \cite{leandro2018} was shown that such a resource it is neither necessary nor sufficient for the mechanism to work, at least when we take in consideration the physical state $\rho_{SR}$ and not the physical vector $\ket{\psi_{SR}}$. However, since it also was shown \cite{tiagomartinelli2019} that the concept of internal coherence is equivalent to the shared asymmetry for $G=U(1)$, the latter being a more general measure, using the theorem we can see that, at least mathematically, a kind of internal entanglement present in the states $\rho_{M}$ remains fundamental to the mechanism because it is equivalent to the concepts of shared asymmetry and internal coherence of the physical state $\rho_{SR}$.
	\item \textbf{Work extracted from internal coherence:} It was shown \cite{kwonjeongjennings2018} that given a $N$-partite system, with non-interacting subsystems, evolving according to the Hamiltonian $H = \sum_{i=1}^N H_i$, where each \textit{i}-th local Hamiltonian $H_i$ has an energy spectrum $\{E_i\}$ with eigenstates $\ket{E_i}$, whose quantum state can be represented as
	\begin{equation}
		\rho = \sum\limits_{\mathbf{E},\mathbf{E'}}\rho_{\mathbf{E},\mathbf{E'}}\ket{\mathbf{E}}\bra{\mathbf{E'}},
	\end{equation}		
	where $\mathbf{E} = (E_1, E_2,..., E_N)$ and $\ket{\mathbf{E}} = \ket{E_1, E_2,..., E_N}$, being $\mathcal{E}_{\mathbf{E}} = \sum\limits_{i=0}^N E_i$ its total energy, the work that can be extracted from the internal coherence of these system, that is the terms with the same total energy $\mathcal{E}_{\mathbf{E}}$, is equal to
	\begin{eqnarray}
		W_{coh} &=& \inf_{\alpha}[F_{\alpha}(\mathcal{D}(\rho)) - F_{\alpha}(\Delta(\rho))] \nonumber \\
		&\leq & F(\mathcal{D}(\rho)) - F(\Delta(\rho)) \nonumber \\
		&=& k_B T[\mathcal{S}(\Delta(\rho)) - \mathcal{S}(\mathcal{D}(\rho))]
	\end{eqnarray}
	where $F(\rho) = \langle E(\rho)\rangle - k_B T\mathcal{S}(\rho)$ is the Helmholtz free energy, $F_{\alpha}(\rho) = k_B T \mathcal{S}_{\alpha}(\rho\|\gamma) - k_B T \log Z$ is the Helmholtz free energy widespread from $\mathcal{S}_{\alpha}(\rho\|\gamma)$, the Rényi divergence \cite{Brandao2015}, $\mathcal{D}(\rho) = \sum_{\mathcal{E}}\Pi_{\mathcal{E}}\rho\Pi_{\mathcal{E}}$ and $\Pi_{\mathcal{E}} = \sum_{\mathbf{E}:\mathcal{E}_\mathbf{E}=\mathcal{E}}\Pi_{\mathbf{E}}$ is the projector in the eigenspace of the total energy $\mathcal{E}$, which corresponds to the charge sector related with the group $U(1)$ action, seen as a group of time-translactions. As noted in Ref. \cite{leandro2018}, bringing this mechanism to the context where $\rho$ is a bipartite system of qubits we can quantify the upper bound of the extracted work using the internal coherence measure, as well
	\begin{equation}
		W(\rho) \leq k_B T C_r(\mathcal{D}(\rho))
	\end{equation}
	It is also possible to extend this relationship to the cases where $\rho$ is a bipartite system of qudits. Using the shared asymmetry notation and applying the theorem, we will find	
	\begin{equation}
		W(\rho) \leq k_B T\mathcal{A}^{(sh)}_{G\otimes G}(\rho) = k_B T\left[\sum_M\Tr(\rho_M^{un})E_R(\rho_M)\right].
	\end{equation}
	This tells that extracting work from internal coherence, at least for bipartite systems, is related with the idea of extracting work from what we called internal entanglement. 
	\item \textbf{Shared asymmetry of states with gap:} As we showed in the last section, an interesting case of shared asymmetry is when we calculate it using a bipartite system of the type: maximally coherent state \textit{versus} a qubit with a gap. The measure value varies from zero to the saturation according to the variation of the gap dimension, from $d-2$ to zero, respectively, according to $d\rightarrow\infty$. We argued that the reason why this happens is linked to the number of blocks in the density matrix, but without at first understanding well what would be behind it. In possession of the theorem we can make an interpretation of this case in relation to the relative entropy of entanglement of the states $\rho_M$ of the charge sectors. If we think about the extreme cases $\ket{\psi_S}=\frac{1}{\sqrt{d}}\sum_{m_S=0}^{d-1} \ket{m_S}$ and $\ket{\psi_R}=\frac{1}{\sqrt{2}}(\ket{0} + \ket{d-1})$ where $\mathcal{A}^{(sh)}_{G\otimes G}(\rho_{SR}) \rightarrow 0$ and $\ket{\psi_S}=\frac{1}{\sqrt{d}}\sum_{m_S=0}^{d-1} \ket{m_S}$ and $\ket{\psi_R}=\frac{1}{\sqrt{2}}(\ket{0} + \ket{1})$ where $\mathcal{A}^{(sh)}_{G\otimes G}(\rho_{SR}) \rightarrow 1$, the first has only one $\rho_M$ entangled (in this case maximally) and the rest are all separable states, the second has $d-1$ entangled states (also maximally) and just two separable states. This leads us to believe that the number of entangled states $\rho_M$ that $\rho_{SR}$ has, in relation to the total number of states $\rho_M$, is directly linked to the value of the shared asymmetry of $\rho_{SR}$.
\end{itemize}    

Besides these interpretations, we can also use the result given in Eq. (\ref{eq_teorema2}) of the theorem to extend the result of the lemma \ref{eq_lema_upper} for more general states. To recap, the lemma says that the shared asymmetry of separable pure states $\rho_{SR}$ always will satisfy the following bounds
\begin{equation}
	0 \leq \mathcal{A}(S:R)\leq \text{min}\{A_G(\rho_S), A_G(\rho_R)\}
\end{equation}
Using the theorem, however, we can show that similar bounds are also valid for the shared asymmetry of more general bipartite states, including mixed states, entangled and mixed entangled, which will allow us to find new states that saturate the bound beyond those presented in the last section. This result we will enunciate as a corollary.   
\begin{coro}
	The shared asymmetry of any state $\rho_{SR}$, for $G=U(1)$, will satisfy the following bounds
	\begin{equation}
		0 \leq \mathcal{A}^{sh}_{G\otimes G}(\rho_{SR}) \leq \min\{\log(\dim\mathcal{H}_S), \log(\dim\mathcal{H}_R)\}
	\end{equation}
\end{coro}
Proof: the lower bound is trivial because once the shared asymmetry is given by a positive linear combination of relative entropies of entanglement, as such entropies are always greater than or equal to zero. Consequently, the same can be said about the shared asymmetry (besides, it had already been demonstrated in Ref. \cite{vaccaro2008}). For the upper bound, we will use the result of the theorem in Eq. (\ref{eq_teorema2}). By this, we can see that shared asymmetry is equal to the weighted average of relative entropies of entanglement for states $\rho_M$. Since the average is always smaller than the maximal value, we have 
\begin{equation}
	\mathcal{A}^{sh}_{G\otimes G}(\rho_{SR}) \leq \max\left\{(E_R(\rho_M))\right\}.
\end{equation} 
Now, to found what is the maximal value of $E_R(\rho_M)$, we just to ask what is the largest support of a state $\rho_M$. The answer is just the largest block that can be found in a state $\mathcal{G}(\rho_{SR})$, for dimensions $\dim\mathcal{H}_S$ and $\dim\mathcal{H}_R$. In other words, this is the number of ways two set of numbers, $\{0,…,\dim\mathcal{H}_S-1\}$ and $\{0,…, \dim\mathcal{H}_R-1\}$, can add to the same number: $\min\{\dim\mathcal{H}_S, \dim\mathcal{H}_R\}$. So, since the largest value of any relative entropy is log the dimension of the system, we get that shared asymmetry is bounded by $\min\{\log(\dim\mathcal{H}_S), \log(\dim\mathcal{H}_R)\}$ as we wanted to demonstrate.

We can separate the consequences of this result in two cases, the trivial and the non-trivial. 
\begin{itemize}
	\item[i)]\textbf{Trivial case}.
	
	We can see that any pure and maximally entangled state $\rho_{SR}$ that is invariant under $\mathcal{G}$ will maximize $\mathcal{A}^{sh}_{G\otimes G}$, since the last term to the right of the above equation will be zero. As a result, we will get $\log(\text{dim}\mathcal{H}_S) = \log(\text{dim}\mathcal{H}_R)$, which is in line with the lemma. This was already expected result, since, working with the internal coherence in the analysis of the Page-Wootters mechanism, in Ref. \cite{leandro2018} it had already been demonstrated that, in the case of Bell-diagonal systems of two qubits, the state $\ket{\psi^{+}} = (\ket{01} + \ket{10})/\sqrt{2}$, which is a maximally entangled state invariant under $\mathcal{G}$, consists in the best possible case. In addition, maximally entangled states had already been identified by Ref. \cite{ReviewQuantumRef} as the best possible case of internal reference frames.   	 
	 \item[ii)]\textbf{Non-trivial case}.
	 
	 This is the case when we have different dimensions $\mbox{dim}\mathcal{H}_S < \mbox{dim}\mathcal{H}_R$, and therefore the states that maximaze the shared asymmetry will be block-diagonal mixed states. 	
	 To see this, lets take a look in one of the several examples from the last section, (equivalent to the example of Eq. (\ref{eq_exemplo_lema}))  
\begin{equation}
    \rho_{SR} = \frac{1}{d(d-1)}\left[\begin{array}{cccccccc}
    A_1 & & & & & & &\\
    &A_2 & & & & & &\\
    & &\ddots & & & & &\\
    & & &A_{d-1} & & & &\\
    & & & &A_{d-1} & & &\\
    & & & & &\ddots & &\\
    & & & & & &A_2 &\\
    & & & & & & &A_1\\
    \end{array}\right].
    \label{eq_exemplo_lema_2}
\end{equation}
Note that of all matrices that constitute the blocks, the one with the largest dimension, $A_{d-1}$, has a dimension equal to the dimension of the Hilbert space of the system $\ket{\psi_S}=\frac{1}{\sqrt{d-2}}\sum_{m_S=0}^{d-3} \ket{m_S}$, being $\mbox{dim}\mathcal{H}_S<\mbox{dim}\mathcal{H}_R$ as proposed. So, using the corollary for these dimensions, $\rho_{SR}^{ideal}$ will saturate the shared asymmetry if it has the following format
\begin{equation}
    \rho_{SR}^{ideal} = \frac{1}{2(d-1)}\left[\begin{array}{cccccccc}
    0 & & & & & & &\\
    &0 & & & & & &\\
    & &\ddots & & & & &\\
    & & &A_{d-1} & & & &\\
    & & & &A_{d-1} & & &\\
    & & & & &\ddots & &\\
    & & & & & &0 &\\
    & & & & & & &0\\
    \end{array}\right].
    \label{eq_exemplo_naotrivial}
\end{equation}
Lets consider that $d=4$. In that case, we have that $\mathcal{A}^{(sh)}_{G\otimes G}(\rho_{SR}^{ideal}) = \log d-1 = \log 3 \approx 1.1$. As for Eq. (\ref{eq_exemplo_lema_2}), calculating the shared asymmetry based on Eq. (\ref{eq_caso_dimdif}), we have that $\mathcal{A}^{(sh)}_{G\otimes G}(\rho_{SR}) = 1/3 + (1/2)\log 3\approx 0.9$. As expected than $\mathcal{A}^{(sh)}_{G\otimes G}(\rho_{SR}^{ideal})>\mathcal{A}^{(sh)}_{G\otimes G}(\rho_{SR})$.
\end{itemize} 

The interesting thing about this last case is precisely the determination of the $\rho_{SR}$ mixed states that saturate the shared asymmetry and, therefore, constitute ideal cases for the functioning of the Page-Wootters mechanism, without the need to consider a system $R$ whose dimension tends to infinity. We called non-trivial not only because they are not so simple to find, unlike the first case, but also because they constitute a new result that can be explored in future works not only within the context of Page-Wootters but also in the area of quantum reference frames in general.

\section{Conclusions and perspectives}

In this work we investigated, from the perspective of quantum reference frame theory, the Page-Wootters mechanism of quantum clocks. Focusing on understanding how well a quantum system can stand as a time reference frame for another quantum system, we concentrated on exploring an informational measure known as shared asymmetry, for the case of $U(1)$ group, that allow us to quantify the necessary resource for the working of the mechanism. We started following the path of the examples presented in Ref. \cite{tiagomartinelli2019} and extended the analysis done there for more general cases, which led us to propose a theorem that relates the shared asymmetry of a bipartite state $\rho_{SR}$ with a sum of the relative entropies of entanglement of what we called internal states $\rho_M$, related to the charge sectors of the Hilbert space $\mathcal{H}_S\otimes\mathcal{H}_R$. These states are by themselves interesting because it is a new class of mixed entangled states similar to the Schmidt correlated states. Besides this, using the theorem we reinterpreted some previous results and defined upper bounds for the shared asymmetry of any bipartite states, which in itself constituted yet another extension of a result from Ref. \cite{tiagomartinelli2019} where the upper bounds were defined only for the case of product states. This last result, in turn, allows us to find a specific type of mixed state that saturates the shared asymmetry consisting of an ideal reference frame, something that had only been shown for systems with a very high dimension or that constitute in a maximally entangled state.  

Based in the results presented in this work, several perspectives for future research have emerged. Bellow we will state some of the main ones that we believe that could be explored making use of the proposals and models presented for other papers in the area of quantum reference frames and quantum information in general.
\begin{itemize}
	\item[•]Extension of the shared asymmetry measure for the case of multipartite systems: 
All the results presented here apply to the case of bipartite systems. However, one question that naturally arises is the possibility of extending of the measure for the case of multipartite systems and gauge symmetries. In order to circumvent the complexity of such a system, one possible path would be use gauge theory techniques to regulate the degrees of freedom that are redundant. Thus, it is possible that $\mathcal{A}_{G\otimes G}^{(sh)}$ can be seen as a measure of the degree of correlation between the system and a gauge field, as proposed in Ref. \cite{tiagomartinelli2019}. Indeed, some works have already used gauge theory to describe quantum process and references, as for example Refs. \cite{gauge2015, gauge2018}.
	\item[•]Study of prohibited operations that can be activated from the resource quantified by the shared asymmetry: One of the most interesting points that is very characteristic of the resource theory approach \cite{teoriasderecurso} is that, once we have access to a state that in a given physical context can be seen as a resource, we can use it to perform operations that in principle would be prohibited by the context. The most common example is the use of shared entangled states between two distant parties in a situation where they are limited to local operations ans classical communication (LOCC) as a resource to perform quantum communication.  	
	Thinking in a similar way about the resource quantified by $\mathcal{A}_{G\otimes G}^{(sh)}$ in the context of a local superselection rule (different of the global superselection rule context that we explored here), we can ask ourselves: which prohibited operation is possible when consuming a quantum state $\rho_{SR}$ for which $\mathcal{A}_{G\otimes G}^{(sh)}(\rho_{SR})\geq 0$? An interesting example that can shed light on the issue is the use of \textit{refbits} (Ref. \cite{refbit2005}) as shared quantum reference units for entanglement activation protocols. States of this type are resources in the same context in which states with shared asymmetry greater than zero also are, that is, under the action of a local superselection rule. However, to activate entanglement in this context, refbits are just useful as the so-called \textit{ebits} and, therefore, if we were to quantify the resource of these states we would have to build a measure that returned the same value for both. In fact, this measure already exist and is known as \textit{Superselection induced variance} (SIV) \cite{schuch2004a, schuch2004b}. However, this does not happen when we apply the shared asymmetry measure to this states: for refbits the result is $1/2$ and for ebits is $1$. This show us that, although $\mathcal{A}_{G\otimes G}^{(sh)}$ is a resource quantifier in the same context which refbits and ebits are resources, the resource quantifier by $\mathcal{A}_{G\otimes G}^{(sh)}$ is not exactly the same as that quantified by SIV. It would, therefore, be interesting to investigate what this resource is and what operations it enables us to do.
	\item[•] Investigation of the properties of type entangled states in Eq. (\ref{eq_estados_B1}):
	Although quite similar to Schmidt's correlated states the mixed tangled states that we present here have the differential that they can also be classified as Werner states \cite{werner1989}. One can, taking advantage of the properties of Werner's states, investigate issues such as the distillation of these states, including the possibility that they are non-distillable, a type known as \textit{bound entangled states}.	
\end{itemize}

\section{Acknowledgments}
The authors would like to thank Leandro R. S. Mendes and Tiago Martinelli for all the helpful discussions and valuable comments. The project was funded by Brazilian funding agencies CNPq (Grant No. 307028/2019-4), FAPESP (Grant No. 2017/03727-0) and the Brazilian National Institute of Science and Technology of Quantum Information (INCT/IQ).

\appendix
\section{Proof of the relations in Eqs. (\ref{dephasing}) and (\ref{fully_dephasing})} 
\label{Apendice_A}

\textit{Proof}: From the point of view of group representation theory, the set of all time translations, given by the unitary operators $U_{\phi} = \{e^{i\phi H};\phi\in[0,2\pi]\}$ where $H = \sum_{m=0}^{d-1}m\ket{m}\bra{m}$, can be seen as an unitary representation of the $U(1)$ compact group (real numbers modulo $2\pi$ under addition). As all unitary representations of a compact group are in general completely reducible and therefore can be decomposed into a discrete quantity of irreducible representations (irreps) we can write \cite{chiribellathesis}
\begin{equation}
	U_{\phi} = \bigoplus_{m\in Q}\bigoplus_{i=1}^{n_m} U^{m,i}_{\phi},
\end{equation}
where $Q$ is the set of the equivalence classes of the irregularities contained in this decomposition, that is, equivalent irreps will be associated with the same $m$ but with a different $i$ index, with $n_m$ being the total number of equivalent irreps which we call \textit{multiplicities}. This decomposition allows us to decompose the Hilbert space $\mathcal{H}$ into orthogonal subspaces that carry the irreps $U^{m,i}_{\phi}$
\begin{equation}
	\mathcal{H} = \bigoplus_{m\in Q}\bigoplus_{i=1}^{n_m}\mathcal{H}_m^i.
	\label{decomp_hilbert}
\end{equation}
It is worth noting that the subspaces $\mathcal{H}_m^i$ are invariant subspaces with respect to the action of the unitary representation $\{U_{\phi}\}$, since $U_{\phi} \ket{\psi} = \ket{\psi'}, \forall \ket{\psi},\ket{\psi'}\in\mathcal{H}_m^i$. In physics, subspaces that carry equivalent irreps, denoted only by $\mathcal{H}_m$ in our case, constitutes what is known as charge sectors. So, in a simple way we can write
\begin{equation}
	\mathcal{H} = \bigoplus_m\mathcal{H}_m.
	\label{decomp_hilbert}
\end{equation}

Since states $\ket{m}$ belonging to the subspaces $\mathcal{H}_m$ are invariant up to a global phase factor in relation to the action of the unitary operator $U_{\phi}$, we can describe the action that operator on an arbitrary state $\ket{\psi}$ (written in the same basis) as
\begin{equation}
	U_{\phi}\ket{\psi} = \sum_m e^{im\phi}\Pi_m\ket{\psi}. 
	\label{eq_projecao_n}
\end{equation}
Replacing this in the equation of the global average, Eq. (\ref{G-twirling_page_wootters}), we have
\begin{eqnarray}
	\mathcal{G}(\ket{\psi}\bra{\psi}) &=& \int_0^{2\pi}\frac{d\phi}{2\pi}U(\phi)\ket{\psi}\bra{\psi}U^{\dagger}(\phi) \nonumber \\
	&=& \int_0^{2\pi}\frac{d\phi}{2\pi}\sum_{m,m'}e^{in\phi}\Pi_m\ket{\psi}\bra{\psi}\Pi_{m'}e^{-im'\phi} \nonumber \\
	&=& \sum_{m,m'}\Pi_m\ket{\psi}\bra{\psi}\Pi_{m'}\left(\int_0^{2\phi}\frac{d\phi}{2\pi}e^{i(m-m')\phi}\right) \nonumber \\
	&=& \sum_m\Pi_m\ket{\psi}\bra{\psi}\Pi_m,
	\label{dephasing_dem}
\end{eqnarray}
that actually is a way of expressing $\mathcal{G}(\cdot)$ that can be used on an arbitrary density operator $\rho$ because it applies to any state $\ket{\psi}$. The equivalence between local average and  fully dephasing, Eq. (\ref{fully_dephasing}), follows directly from that.

In the case of the global average acting on a bipartite spate $\rho_{SR}$ we have that the Hamiltonian is $H = H^S\otimes\mathds{1}^R + \mathds{1}^S\otimes H^R$ where $H^{\alpha} = \sum_{m=0}^{d-1}m\ket{m}\bra{m}$, $\alpha = S,R$, so the action of the unitary operator $U_{\phi} = e^{i\phi(H^S\otimes\mathds{1}^R + \mathds{1}^S\otimes H^R)}$ on an arbitrary state $\ket{\psi_{SR}}$ can be described by
\begin{eqnarray}
	U_{\phi}\ket{\psi_{SR}} &=& \sum_{M}\sum_{M=m_S+m_R} e^{i(m_S+m_R)\phi}\Pi_{m_S}\otimes\Pi_{m_R}\ket{\psi_{SR}} \nonumber \\
	&=& \sum_{M} e^{iM\phi}\Pi_M\ket{\psi_{SR}}, 
	\label{eq_projecao_n}
\end{eqnarray}
where $\Pi_M = \sum_{M=m_S+m_R}\Pi_{m_S}\otimes\Pi_{m_R}$ is the projection onto the charge sectors $\mathcal{H}_M$ of the Hilbert space $\mathcal{H}_S\otimes\mathcal{H}_R$. The extension to the multipartite case is straightforward.

Its worth mention that for general finite or compact groups $G$, with unitary representations $U_g$, the  charge sectors $\mathcal{H}_q$ can be further decomposed into a \textit{virtual} tensor product,
\begin{equation}
	\mathcal{H}_q = \mathcal{M}_q\otimes\mathcal{N}_q,
\end{equation} 
where $\mathcal{M}_q$ is a subsystem carrying an irrep $U_q$ and $\mathcal{N}_q$ is a subsystem carrying a trivial representation of $G$. These subsystem are \textit{virtual} \cite{zanardi2001} since they do not describe individual physical states. In literature, they are often called gauge and multiplicity space or color space and flavor space, respectively. For our case the group is and Abelian group so the irreps are one-dimension and therefore the subspaces $\mathcal{M}_q$ are trivial and can be disregarded. However, in cases of non-Abelian groups such as $SU(2)$ in which this does not happen, making this separation between subsystems is important because operation $\mathcal{G}$ will act differently in each of them, and can no longer be written in a simplified way just like Eq. (\ref{dephasing_dem}). For these cases the action of $\mathcal{G}$ in an arbitrary density operator $\rho$ is given by
\begin{equation}
	\mathcal{G}[\rho] = \sum_q(\mathcal{D}_{\mathcal{M}_q}\otimes\mathcal{I}_{\mathcal{N}_q})[\Pi_q\rho\Pi_q]
	\label{twirl_geral}
\end{equation}
where $\Pi_q$ are the projectors in the charge sectors $\mathcal{H}_q$, $\mathcal{D}_{\mathcal{M}}$ is a trace-preserving operation that takes every operator on the Hilbert space $\mathcal{M}_q$ to a constant times the identity operator on that space, and $\mathcal{I}_{\mathcal{N}}$ denotes the identity map over operators in the space $\mathcal{N}$. The proof of this result can be found in Ref \cite{ReviewQuantumRef}.

\section{Relative entropy of entanglement for a class of mixed states} 
\label{Apendice_B} 

In this appendix we will demonstrate that for a mixed state of the type
\begin{equation}
	\rho = \sum\limits_{{n_1} = 0}^N\sum\limits_{{n_2} = 0}^N c_{n_1,n_2}\ket{n_1,N-n_1}\bra{n_2,N-n_2},
	\label{eq_estado}
\end{equation} 
the separable state that minimizes the relative entropy, $E_R(\rho)$, will be
\begin{equation}
	\sigma^{*} = \sum\limits_{{n_1} = 0}^N c_{n_1}\ket{n_1,N-n_1}\bra{n_1,N-n_1}
	\label{eq_estado_sep}.
\end{equation}
Thus, $E_{R}(\rho) = \mathcal{S}(\sigma) - \mathcal{S}(\rho)$, where $S(\rho)=-\Tr\rho\ln\rho$ is the von Neumann entropy.
This demonstration follows directly from the one made in Ref. \cite{wuzhang2000} for mixed states of the type $\sum\limits_{n_1,n_2}a_{n_1,n_2}\ket{\phi_{n_1},\psi_{n_1}}\bra{\phi_{n_2},\psi_{n_2}}$, known as Schmidt correlate states. 

As we already have a guess for the separable state that minimizes the relative entropy of  (\ref{eq_estado}), following the Vedral-Plenio theorem \cite{vedralplenio1998}, we need to show that the gradient $\frac{d}{dx}\mathcal{S}(\rho\parallel(1-x)\sigma^{*} + x\sigma)\mid_{x=0}$ for all $\sigma \in \text{SEP}$, where $\text{SEP}$ is the set of all non-entangled states, is not negative. If this does not happen for a given state $\sigma^{*}$ that means that it is not a minimum of the function $f(x,\sigma) = \mathcal{S}(\rho\parallel(1-x)\sigma^{*} + x\sigma)$ and, therefore, the guess is wrong.

Using the identity $\ln A = \int_{0}^{\infty}[(At - 1)/(A + t)]dt/(1 + t^{2})$, we can write the gradient as
\begin{equation}
	\frac{\delta f}{\delta x}(0,\sigma) = 1 - \int_{0}^{\infty}\Tr[(\sigma^{*} + t)^{-1}\rho(\sigma^{*} + t)^{-1}\sigma]dt
	\label{eq_prova}.
\end{equation}
Replacing (\ref{eq_estado}) and (\ref{eq_estado_sep}) in $(\sigma^{*} + t)^{-1}\rho(\sigma^{*} + t)^{-1}$, we will have
\begin{eqnarray}
	&(&\sigma^{*} + t)^{-1}\rho(\sigma^{*} + t)^{-1} \nonumber \\
	&=& \left(\sum\limits_{n_1}c_{n_1,n_1}\ket{n_1,N-n_1}\bra{n_1,N-n_1} + t\right)^{-1} \nonumber \\
	&\times & \sum\limits_{n_2,n_3}c_{n_2,n_3}\ket{n_2,N-n_2}\bra{n_3,N-n_3} \nonumber \\
	&\times & \left(\sum\limits_{n_4}c_{n_4,n_4}\ket{n_4,N-n_4}\bra{n_4,N-n_4} + t\right)^{-1} \nonumber \\
	&=& \sum\limits_{n_1,n_2,n_3,n_4}(c_{n_1,n_1} + t)^{-1}c_{n_2,n_3}(c_{n_4,n_4} + t)^{-1} \nonumber \\
	&\times & \ket{n_1,N-n_1}\bra{n_1,N-n_1}\ket{n_2,N-n_2} \nonumber \\
	&\times & \bra{n_3,N-n_3}\ket{n_4,N-n_4}\bra{n_4,N-n_4}.
\end{eqnarray} 
In the last two lines it is noted that the two brakets will generate two Dirac deltas, $\delta_{n_1,n_2}$ and $\delta_{n_3,n_4}$, so for the equation to be different from $0$ we will make $n_1 = n_2 = n$ and $n_3 = n_4 = n^{'}$,
\begin{align}
	&(\sigma^{*} + t)^{-1}\rho(\sigma^{*} + t)^{-1}= \notag \\
	&\sum\limits_{n,n^{'}}(c_{n,n} + t)^{-1}c_{n,n^{'}}(c_{n^{'},n^{'}} + t)^{-1}\ket{n,N-n}\bra{n^{'}, N-n^{'}}.
\end{align}		

Let's $g(n,n^{'})\equiv c_{n,n^{'}}\int_{0}^{\infty}(c_{n,n} + t)^{-1}(c_{n^{'},n^{'}} + t)^{-1}dt$, obviously $g(n,n)=1$ and for $n\neq n^{'}$,
\begin{equation}
	g(n,n^{'}) = c_{n,n^{'}}\;\frac{\log c_{n,n} - \log c_{n,n^{'}}}{c_{n,n} - c_{n^{'},n^{'}}}.
\end{equation} 
Now we will show that $|g(n,n^{'})|\leq 1$. As the Vedral-Plenio theorem proved that 
\begin{equation}
	0\leq\sqrt{c_{n,n}c_{n^{'}},c_{n^{'}}}\;\frac{\log c_{n,n} - \log c_{n^{'},n^{'}}}{c_{n,n} - c_{n^{'},n^{'}}}\leq 1,
\end{equation}
we just need to prove that $|c_{n,n^{'}}|\leq\sqrt{c_{n,n}c_{n^{'},n^{'}}}$. To do so, let $\ket{\psi} = a\ket{n,N-n} + b\ket{n^{'},N-n^{'}}$, where $a,b \in\mathbb{C}$, we will have
\begin{equation}
	\bra{\psi}\rho\ket{\psi}\geq 0,
\end{equation}
\begin{equation}
	|a|^{2}c_{n,n} + |b|^{2}c_{n^{'},n^{'}} + a^{*}bc_{n,n^{'}} + ab^{*}c_{n^{'},n}\geq 0.
\end{equation}
This last inequality can be written as a density matrix,
\begin{equation}
	\left[\begin{array}{cc}
    |a|^{2}c_{n,n} - \lambda & a^{*}b\;c_{n,n^{'}} \\
    ab^{*}c_{n^{'},n} & |b|^{2}c_{n^{'},n^{'}} - \lambda 
	\end{array}\right].\\
\end{equation}  
Diagonalizing it and considering that $\lambda \geq 0$, we will have
\begin{equation}
	(|a|^{2}c_{n,n} + |b|^{2}c_{n^{'},n^{'}})^{2} \geq  (\sqrt{\bigtriangleup})^{2},
\end{equation}
where $\bigtriangleup = (|a|^{2}c_{n,n} - |b|^{2}c_{n^{'},n^{'}})^{2} + 4|ab|^{2}|c_{n,n^{'}}|^{2}$. 
After some simplifications it can be shown that
\begin{equation}
	\sqrt{c_{n,n}c_{n^{'},n^{'}}}\geq |c_{n,n^{'}}|,
\end{equation}
as we wanted to demonstrate. Therefore, $|g(n,n^{'})|\leq 1$.

Now be $\sigma\equiv\ket{\alpha}\bra{\alpha}\otimes\ket{\beta}\bra{\beta}$, with $\ket{\alpha} = \sum\limits_{n=0}^N a_n\ket{n}$ and $\ket{\beta}=\sum\limits_{n=0}^N b_n\ket{N-n}$ are normalized vectors. So, going back to the Eq. (\ref{eq_prova}), we can write
\begin{eqnarray}
	&\frac{\delta f}{\delta x}&(0,\sigma) - 1 \nonumber \\ 
	&=& -\Tr\left[\sum\limits_{n_1,n_2,n_3,n_4,n_5,n_6}g(n_1,n_2)\ket{n_1,N-n_1}\bra{n_2,N-n_2}\right. \nonumber \\
	&\times & \left. a_{n_3}\overline{a_{n_5}}b_{n_4}\overline{b_{n_6}}\ket{n_3}\bra{n_5}\otimes\ket{N-n_4}\bra{N-n_5}\right] \nonumber \\
	&=& -\sum\limits_{n_1,n_2}g(n_1,n_2)\;a_{n_2}\;b_{n_2}\;\overline{a_{n_1}}\;\overline{b_{n_1}}.
\end{eqnarray}
Therefore, using the Cauchy-Schwarz inequality
\begin{align}
	\left|\frac{\delta f}{\delta x}(0,\sigma) - 1\right| &\leq  \sum\limits_{n_1,n_2}|g(n_1,n_2)||			a_{n_2}||b_{n_2}||\overline{a_{n_1}}||\overline{b_{n_1}}| \nonumber \\
	&\leq  \sum\limits_{n_1,n_2}|a_{n_2}||b_{n_2}||\overline{a_{n_1}}||\overline{b_{n_1}}| \nonumber \\
	&= \left(\sum\limits_{n}|a_n||b_n|\right)^{2}\leq\sum\limits_{n}|a_n|^{2}|b_n|^{2} = 1,
\end{align}
we come to
\begin{equation}
	\frac{\delta f}{\delta x}(0,\ket{\alpha,\beta}\bra{\alpha,\beta})\geq 0.
\end{equation}
Since all non-entangled state can be written as $\sigma = \sum\limits_i r_i\ket{\alpha^i,\beta^i}\bra{\alpha^i,\beta^i}$, we have that
\begin{equation}
	\frac{\delta f}{\delta x}(0,\sigma) = \sum\limits_i r_i \frac{\delta f}{\delta x}(0,\ket{\alpha^i,\beta^i}\bra{\alpha^i,\beta^i})\geq 0.
\end{equation}
Thus, we shown that, in fact, the gradient $\frac{d}{dx}\mathcal{S}(\rho\parallel(1-x)\sigma^{*} + x\sigma)\mid_{x=0}$ is non-negative, which indicates that $\sigma^{*}$ is the separable state that minimizes relative entropy. To confirm, it remains only to show that $\mathcal{S}(\rho\parallel\sigma)\geq \mathcal{S}(\rho\parallel\sigma^{*}), \forall\sigma\in \text{SEP}$. The proof will be make by contradiction: suppose that $\mathcal{S}(\rho\parallel\sigma)<\mathcal{S}(\rho\parallel\sigma^{*})$, for some $\sigma\in \text{SEP}$, so for $0 < x \leq 1$,
\begin{eqnarray}
	f(x,\sigma) &=& \mathcal{S}(\rho\parallel(1-x)\sigma^{*} + x\sigma) \nonumber \\
	&\leq & (1-x)\mathcal{S}(\rho\parallel\sigma^{*}) + x\mathcal{S}(\rho\parallel\sigma) \nonumber \\
	&=& (1-x)f(0,\sigma) + xf(1,\sigma), 
\end{eqnarray}
thus,
\begin{eqnarray}
	f(x,\sigma) &\leq & f(0,\sigma) - xf(0,\sigma) + xf(1,\sigma) \nonumber \\
	f(x,\sigma) - f(0,\sigma) &\leq & x(f(1,\sigma) - f(0,\sigma)) \nonumber \\
	\frac{f(x,\sigma) - f(0,\sigma)}{x} &\leq & f(1,\rho) - f(0,\rho) < 0.
\end{eqnarray}
This contradicts the fact that $\frac{\delta f}{\delta x}(0,\sigma)>0$ in the limit of $x\rightarrow 0$.
So, $\mathcal{S}(\rho\parallel\sigma)\geq\mathcal{S}(\rho\parallel\sigma^{*})$ as we wanted to demonstrate and $\sigma^{*} = \sum\limits_{n=0}^k c_{n,n}\ket{n,N-n}\bra{n,N-n}$ minimizes relative entropy.
 
 \vspace*{\fill}


%

\end{document}